\documentclass[aps,showkeys]{revtex4}

\usepackage{epsfig}
\usepackage{eepic}
\usepackage{latexsym}
\usepackage{rotating}
\usepackage{lscape}
\usepackage{graphics}
\usepackage{dcolumn}
\usepackage{bm}
\newcommand{\beq}{\begin{equation}}
\newcommand{\eeq}{\end{equation}}
\newcommand{\bd}{\begin{displaymath}}
\newcommand{\ed}{\end{displaymath}}

\newcommand{\bei}{\begin{itemize}}
\newcommand{\eei}{\end{itemize}}

\newcommand{\bee}{\begin{enumerate}}
\newcommand{\eee}{\end{enumerate}}

\begin{document}

\noindent
\title{Computation of the electron beam quality $k_{{\rm Q,Q}_0}$ factors for the NE2571, NE2571A and NE2581A thimble ionization chambers using PENELOPE}

\author{F. Erazo$^{1,2}$, L. Brualla$^3$ and A.M. Lallena$^4$ \\
{\small {\it
$^1$Instituto del C\'ancer - SOLCA, Cuenca, Ecuador. \\
$^2$Escuela de Tecnolog\'{\i}a M\'edica, Facultad de Medicina, Universidad de Cuenca, Cuenca, Ecuador. \\
$^3$NCTeam, Strahlenklinik, Universit\"atsklinikum Essen, Hufelandstra\ss e 55, D-45122 Essen, Germany.\\
$^4$Departamento de F\'isica At\'omica, Molecular y Nuclear,
Universidad de Granada, E-18071 Granada, Spain.
}}}

\date{\today}

\bigskip

\begin{abstract}
\noindent
The quality correction factor $k_{Q,Q_{0}}$ for electron beams was calculated for three thimble ionization chambers, namely, NE2571, NE2571A and NE2581A. The Monte Carlo code PENELOPE was used to estimate the overall correction factor $f_{\rm c,Q}$ of these chambers for electron beams with nominal energies ranging between 6 and 22 MeV, corresponding to a Varian Clinac 2100 C/D. A $^{60}$Co beam was used as reference quality ${\rm Q}_0$. Also eight monoenergetic electron beams reproducing the quality index $R_{50}$ of the Clinac beams were considered. The $k_{Q,Q_{0}}$ factors were calculated as the ratio between $f_{\rm c,Q}$ and $f_{{\rm c,Q}_0}$. Those obtained for the NE2571 ionization chamber show a nice agreement with those calculated by Muir and Rogers with EGSnrc. As it occurred to other ionization chambers analyzed in previous works, the $k_{{\rm Q,Q}_0}$ factors found for the monoenergetic beams are larger (smaller) than those corresponding to the Clinac beams at low (high) $R_{50}$ values, the differences being slightly above $0.5\%$. Finally, the $k_{{\rm Q,Q}_0}$ factors obtained in the case of the NE2571A chamber are systematically $\sim 0.5\%$ below those of its predecessor chamber, the NE2571.
\end{abstract}

\keywords{$k_{Q,Q_{0}}$ correction factors, electron beams, NE2571, NE2571A, NE2581A, PENELOPE}

\maketitle

\section*{Introduction}

Historically plane-parallel ionization chambers had been recommended for calibrating low energy clinical electron beams in reference dosimetry \cite{Almond99,IAEA00}. However, Muir {\it et al.} \cite{Muir12} quoted long-term instabilities of the calibration coefficients for this type of chambers. As a consequence, cylindrical ionization chambers, that were not previously considered for reference dosimetry of electron beams with energies below $10\,$MeV, turned to be of interest for that purpose.

Muir and Rogers \cite{Muir13} investigated the capabilities of this type of chambers for the calibration of clinical electron beams by using the Monte Carlo code EGSnrc \cite{Kawrakow00,Kawrakow07}. Specifically, they calculated and analyzed the beam quality conversion factors for the PTW Roos plane-parallel chamber and the Farmer-type NE2571 thimble chamber considering different electron beams with nominal energies ranging between 4 and $22\,$MeV. 

The beam quality correction factor is given by \cite{Sempau04,Capote04}: 
\begin{equation}
k_{{\rm Q,Q}_0} \, = \, \frac{f_{\rm c,Q}}{f_{{\rm c,Q}_0}} \, ,
\label{eq:kQQ0}
\end{equation}
with
\begin{equation}
f_{\rm c,Q}\, = \, \frac{D_{\rm w,Q}}{D_{\rm c,Q}}  \, . 
\label{eq:fcQ}
\end{equation}
and
\begin{equation}
f_{{\rm c,Q}_0}\, = \, \frac{D_{{\rm w,Q}_0}}{D_{{\rm c,Q}_0}}  \, . 
\label{eq:fcQ0}
\end{equation}
the overall perturbation factors for the beam qualities Q and Q$_0$, the latter being the reference one. Thus, the factor $f_{\rm c,Q}$ relates directly the absorbed dose at the reference point in water, $D_{\rm w,Q}$, with the absorbed dose in the air cavity of the actual detector, $D_{\rm c,Q}$ used in the calibration process and situated at the same point. In other words, $f_{\rm c,Q}$ takes into account the whole effect linked to the presence of the detector in the water medium where measurements are carried out. The value of $f_{\rm c,Q}$ may be obtained by Monte Carlo simulation provided an accurate geometrical description of the detector is available.

Equation (\ref{eq:kQQ0}) was employed to evaluate the beam quality correction factor corresponding to various chambers irradiated with both photon and electron beams and using different Monte Carlo codes \cite{Muir13,Zink08,Gonzalez09,Muir10,Zink12,Erazo13,Erazo14,Erazo16,Reis16}. 
The aim of the present work is to use the Monte Carlo code PENELOPE \cite{Salvat14} to estimate $k_{{\rm Q,Q}_0}$, in case of electrons beams, for three ionization chambers: the NE2571, permitting a comparison with the results of Muir and Rogers quoted above, and the NE2571A and the NE2581A thimble ionization chambers, both manufactured by QADOS (Sandhurst, UK). Eight electron beams from a Varian Clinac 2100 C/D (Varian Medical Systems, Palo Alto, USA), with nominal energies between 6 and 22 MeV, and also eight monoenergetic electron beams providing the same quality as those from the Clinac, were simulated as irradiation sources. Differences between the results obtained for the three ionization chambers were also investigated.

\section*{Material and methods}

\subsection*{Ionization chambers}

As said above, three cylindrical thimble type ionization chambers were considered in this work. The first
one was the NE2571 Farmer-type chamber whose geometry was obtained from the works  of  Aird and Farmer \cite{Aird72} and Wulff {\it et al.} \cite{Wulff08}. The chamber cavity has a diameter of 0.64 cm and a length of 2.40 cm. The central electrode, made of aluminium, has a diameter of 0.1 cm and the wall is of graphite, with a thickness of 0.04 cm. The electrode and the wall are separated by PTFE (polytetrafluoroethylene) that acts as insulator. The chamber model includes a sleeve of 0.10 cm made of  PMMA (polymethyl methacrylate). The nominal volume of the chamber is 0.6 cm$^{3}$. The geometrical description used in this work was the same as that considered in \cite{Erazo13} to calculate the $k_{{\rm Q,Q}_0}$ factors for photon beams that showed a nice agreement with those determined in \cite{Muir10,Wulff08}.

Details of the other two chambers considered, the NE2571A and the NE2581A, were provided by the manufacturer and are described in the work by Erazo and Lallena \cite{Erazo16}. These are Farmer-type chambers with sensitive volumes of 0.69 and $0.56\,$cm$^3$, respectively. 
Their external diameter is $0.86\,$cm and have a length (including the connecting cable) of about $9\,$cm. The diameter of the air cavity of both chambers is $0.63\,$cm, their length is $2.41\,$cm and the wall thickness is $0.036\,$cm. In the NE2571A chamber, the thimble is made of graphite and the inner electrode has a diameter of $0.1\,$cm and is made of Al. In the NE2581A chamber, both the thimble and the electrode (with a diameter of $0.3\,$cm) are made of Shonka A-150. An insulator of PTFE between the electrode and the wall is present in both chambers.

\subsection*{Monte Carlo simulations}

Using the overall perturbation factors calculated with equations (\ref{eq:fcQ}) and (\ref{eq:fcQ0}), the beam quality correction factors $k_{{\rm Q,Q}_0}$ were obtained for these chambers using equation (\ref{eq:kQQ0}). To estimate the absorbed doses at the medium, $D_{\rm w,Q}$, and the detector, $D_{\rm c,Q}$, a series of Monte Carlo simulations were run with the main program {\sc penEasy} \cite{Sempau09}. This code is a main steering program that uses  the PENELOPE system \cite{Salvat14}, a Monte Carlo general-purpose code designed for an accurate description of the coupled transport of photons, electrons and positrons in matter, with very good achievements at the material interfaces and for low energies \cite{Sempau03,Faddegon08,Faddegon09,Salvat09,Vilches09}. The code has also been benchmarked for simulations of ionization chamber responses in calculations similar to those done in the present work \cite{Sempau06,Yi06}. 

In PENELOPE the photon transport is done according to the detailed method, describing the interactions they suffer in sequential order, while the transport of electrons and positrons is simulated within a mixed scheme that involves both hard and soft collisions. The first ones are simulated in a detailed way; interactions of the second type occurring between two consecutive hard collisions are described by means of a multiple scattering theory. The selection of hard or soft interactions is done by fixing threshold values of the polar deflection angle and the energy loss. Eight parameters permit to control the electron and positron simulation: $C_1$ (the average angular deflection produced by all soft interactions occurred along a path length equal to the mean free path between consecutive hard elastic events), $C_2$ (the maximum average fractional energy loss between consecutive hard elastic events), $W_{\rm CC}$ and $W_{\rm CR}$ (the cut-off energy values for hard inelastic collisions and bremsstrahlung emission, respectively), $E_{\rm ABS}(\gamma)$, $E_{\rm ABS}({\rm e}^-)$ and $E_{\rm ABS}({\rm e}^+)$ (the absorption energies below which the corresponding particle is stopped) and $s_{\rm max}$ (the maximum length permitted to a simulation step). These parameters must be chosen by the user according to the specific calculation to be done.

The simulations performed in the present work were carried out in a way similar to those of previous works \cite{Erazo13,Erazo14,Erazo16}, following the prescription of Sempau and Andreo \cite{Sempau06} for the optimal values of the PENELOPE tracking parameters to simulate ionization chambers. Specifically, we used $C_1=C_2=0.02$ for all materials of the ionization chamber and a water volume extending $2\,$cm around it. For the rest of the materials in the simulation geometry we fixed $C_1=C_2=0.1$. For the absorption energies we considered $E_{\rm ABS}({\rm e}^-)=E_{\rm ABS}({\rm e}^+)=0.01\, E_{\rm max}$,
$E_{\rm ABS}(\gamma)=0.001\, E_{\rm max}$, with $E_{\rm max}$ the energy of the primary electrons at the source. Finally, the parameter $s_{\rm max}$ was fixed to one tenth of the thickness of the corresponding material.

\subsection*{Simulation geometry}

The simulation geometry was that recommended in the TRS-398 protocol (IAEA 2000) for the measuring conditions that includes a water phantom of $50\times 50\times 50$~cm$^3$ situated at a source-to-surface distance of 100 cm. An irradiation field of $10\times 10$~cm$^2$ was assumed and the reference depth was determined according to $z_{\rm ref}=0.6\,R_{50}-0.1$~cm.
 
The absorbed dose scored within the chamber active volume provided us with the doses $D_{\rm c,Q}$. The scoring voxel considered to estimate the dose absorbed in water, $D_{\rm w,Q}$, was similar to that used by other authors in previous calculations \cite{Muir13,Zink08,Zink12,Erazo14,Zink09}. It consisted of a cylinder with a radius of 1 cm and a height of 0.025 cm centered in the beam axis and situated at the reference depth.  

\subsection*{Radiation sources}

Eight electron beams corresponding to a Varian Clinac 2100 C/D with nominal electron energies of 6, 9, 12, 15, 16, 18, 20 and 22 MeV were considered as radiation sources. The linac geometries were generated with the code {\sc penEasyLinac} \cite{Brualla09,Sempau11}, a code that has been tested in previous works \cite{Brualla09,Sempau11,Isambert10,Brualla12}.
 
These beams were simulated by, first, generating phase-space files (PSFs) at a surface normal to the beam axis situated at the entrance of the water phantom. In a second step the PSFs were used as sources of particles that were emitted towards the phantom.

To estimate the effects of the whole Clinac geometries, eight monoenergetic beams were also considered. These monoenergetic beams were tuned in such a way that the $R_{50}$ values of the corresponding Clinac beams were reproduced. This was done by calculating the percentage depth dose curves in the water phantom described above. In these simulations, however, we used cylindrical scoring voxels with a height of $0.2\,$cm and a radial thickness of $0.5\,$cm, concentrical to the beam axis. The monoenergetic beams were collimated mathematically to conform the radiation fields; that is, the transport of the particles arriving at the phantom entrance outside the defined square field was discontinued.

\begin{table}[!th]
\caption{\label{tab:R50} Energy $E_{\rm ini}$ of the primary electron beams used in the simulation of the Varian Clinac 2100 C/D and values of $R_{50}$ found for each nominal energy. Also the $R_{50}$ values obtained for the corresponding monoenergetic beams and their tuned initial energies are given. The maximum uncertainty in the estimations of the $R_{50}$ values is 0.1\% (with a coverage factor $k=1$).
}  
\begin{center}
\begin{tabular}{cccccc}
\hline\hline
& \multicolumn{2}{c}{Clinac} &~~~& \multicolumn{2}{c}{monoenergetic}  \\ \cline{2-3} \cline{5-6}
nominal energy  & $E_{\rm ini}$ & $R_{50}$ & & $E_{\rm tuned}$ & $R_{50}$    \\ 
{[MeV]} & [MeV] & [cm] & &  [MeV] & [cm]  \\ 
\hline
6 & 7.27  & 2.54 & & 6.47 & 2.53 \\
9& 10.21 & 3.73 & & 9.22 & 3.73  \\ 
12& 13.37  & 4.98 & & 12.11& 4.98 \\ 
15& 16.61  & 6.27 & & 15.15& 6.27  \\ 
16& 17.80  & 6.70 & & 16.19  & 6.72  \\ 
18& 19.97  & 7.50 & &  18.12& 7.51  \\ 
20& 22.23  & 8.24 &  & 19.96 & 8.25  \\ 
22& 24.46  & 9.01 & & 21.96 & 9.03 \\ 
\hline\hline
\end{tabular}
\end{center}
\end{table}

Table \ref{tab:R50} summarizes the information concerning the beams considered in the simulations. The energies $E_{\rm ini}$ of the primary electron beams used in the simulation of the Varian Clinac 2100 C/D were obtained by tuning the corresponding simulations performed with PRIMO/{\sc penEasyLinac} \cite{Brualla09,Sempau11,PRIMO} in order to reproduce the depth doses of the corresponding golden beam data provided by Varian for this linac. In table \ref{tab:R50}, also the $R_{50}$ values corresponding to the different Clinac beams as well as those obtained for the tuned monoenergetic beams are included. The relative differences between the values corresponding to the full linac models and those found after tuning the monoenergetic beams are below 0.3\%. 

Potential functions of the type
\begin{equation}
f_{\rm chamber}^{\rm beam\,type}\, = \, a \, + \, b \, x^c
\label{eq:fit}
\end{equation}
were fitted to the values of $k_{{\rm Q,Q}_0}$ vs. ${R_{50}}$ (in cm) obtained in our simulations for the three chambers and both for Clinac and monoenergetic beams. 

For the reference quality Q$_0$, a $^{60}$Co gamma beam was considered. The corresponding simulations were done in a way similar to that used for the monoenergetic beams with the emitted photons having the spectrum quoted by Mora and Maio \cite{Mora99}.

\section*{Results and discussion}

The $f_{{\rm c,Q}_0}$ values obtained for the $^{60}$Co beam were $1.1130\pm 0.0015$ for the NE2571 chamber, $1.1174\pm 0.0015$ for the NE2571A and $1.1430\pm 0.0016$ for the NE2581A.

\begin{figure}[!b]
\centering
\includegraphics[width=6cm]{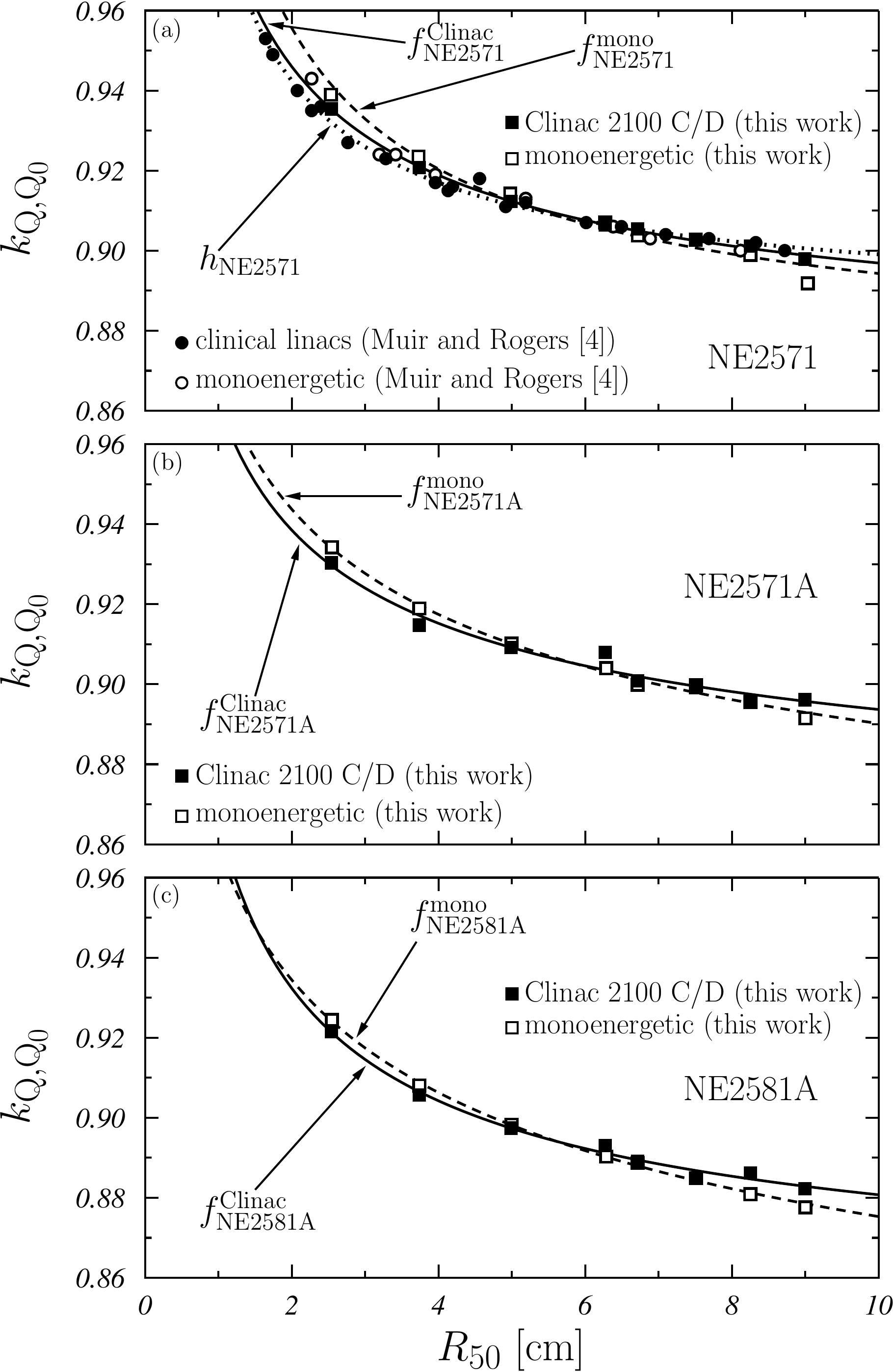}
\caption{Values of $k_{{\rm Q,Q}_0}$, as a function of $R_{50}$, obtained for the three ionization chambers studied when the Clinac beams (solid squares) and the monoenergetic beams reproducing their $R_{50}$ values (open squares) are considered as radiation sources. The results are compared to those quoted by Muir and Rogers \cite{Muir13} for both monoenergetic beams (open circles) and clinical beams (solid circles). Solid and dashed curves are the fits of the function defined in equation (\ref{eq:fit}) to the results we obtained for the Clinac and the monoenergetic beams, respectively. The corresponding fitting parameters are given in table \ref{tab:param}. The dotted curve in panel (a) represents the fit in equation (\ref{eq:fitMR}) given in reference \cite{Muir13}.
\label{fig:UNO}
}
\end{figure}

\begin{table}[!th]
\caption{\label{tab:DOS} 
$k_{{\rm Q,Q}_0}$ values obtained in our simulations for the three ionization chambers studied. The results found for the Clinac and monoenergetic beams are shown. The uncertainty (with a coverage factor $k=1$) is given between parentheses; thus, 0.9390(15) means $0.9390 \pm 0.0015$.}
\begin{center}
{\scriptsize
\begin{tabular}{cccccc}
\hline\hline
& \multicolumn{2}{c}{Clinac} &~~~& \multicolumn{2}{c}{monoenergetic}  \\ \cline{2-3} \cline{5-6}
chamber  & $R_{50}$ [cm]Ê& $k_{{\rm Q,Q}_0}$ & & $R_{50}$ [cm] & $k_{{\rm Q,Q}_0}$ \\ 
\hline
 NE2571
 &  2.54 & 0.9354(10) &&  2.53 & 0.9390(15) \\
 &  3.73 & 0.9208(11) &&  3.73 & 0.9235(16) \\
 &  4.98 & 0.9123(10) &&  4.98 & 0.9142(17) \\
 &  6.27 & 0.9065(11) &&  6.27 & 0.9070(17) \\
 &  6.70 & 0.9054(11) &&  6.72 & 0.9038(17) \\
 &  7.50 & 0.9027(12) &&  7.51 & 0.9027(17) \\
 &  8.24 & 0.9012(11) &&  8.25 & 0.8990(17) \\
 &  9.01 & 0.8979(12) &&  9.03 & 0.8918(18) \\
 \hline
 NE2571A
 &  2.54 & 0.9304(11) &&  2.53 & 0.9342( 8) \\
 &  3.73 & 0.9148(12) &&  3.73 & 0.9189( 8) \\
 &  4.98 & 0.9092(11) &&  4.98 & 0.9102( 8) \\
 &  6.27 & 0.9079(12) &&  6.27 & 0.9040( 9) \\
 &  6.70 & 0.9008(12) &&  6.72 & 0.8999( 9) \\
 &  7.50 & 0.8992(12) &&  7.51 & 0.8998( 9) \\
 &  8.24 & 0.8956(12) &&  8.25 & 0.8955( 9) \\
 &  9.01 & 0.8961(12) &&  9.03 & 0.8915( 9) \\
 \hline
 NE2581A
 &  2.54 & 0.9216(11) &&  2.53 & 0.9244( 6) \\
 &  3.73 & 0.9057(12) &&  3.73 & 0.9081( 6) \\
 &  4.98 & 0.8974(11) &&  4.98 & 0.8982( 7) \\
 &  6.27 & 0.8931(11) &&  6.27 & 0.8903( 7) \\
 &  6.70 & 0.8891(12) &&  6.72 & 0.8887( 7) \\
 &  7.50 & 0.8849(12) &&  7.51 & 0.8849( 7) \\
 &  8.24 & 0.8862(12) &&  8.25 & 0.8809( 8) \\
 &  9.01 & 0.8822(12) &&  9.03 & 0.8776( 9) \\
 \hline\hline
\end{tabular}
}
\end{center}
\end{table}

Figure \ref{fig:UNO} and table \ref{tab:DOS} show the $k_{{\rm Q,Q}_0}$ calculated for the three ionization chambers studied in the conditions described above. In the figure solid and open squares represent the values we have obtained for the Clinac and the monoenergetic beams, respectively. The solid and dashed curves are the corresponding fitting functions as defined in equation (\ref{eq:fit}). The values of the fitting parameters are shown in table \ref{tab:param}.

\begin{table}[!b]
\caption{\label{tab:param} 
Fitting parameters of the function $f_{\rm chamber}^{\rm beam\,type}$, defined in equation (\ref{eq:fit}), for the values $k_{{\rm Q,Q}_0}$ vs. ${R_{50}}/{\rm cm}$ obtained in our simulations for the three ionization chambers studied. The uncertainty (with a coverage factor $k=1$) is given between parentheses; thus, 0.120(2) means $0.120\pm 0.002$.}
\begin{center}
\begin{tabular}{cccc}
\hline\hline
fitting function & $a$ & $b$ & $c$  \\ \hline
$f_{\rm NE2571}^{\rm Clinac}$ & 0.868(7) &  0.120(2) & 0.62(9) \\
$f_{\rm NE2571}^{\rm mono}$ & 0.867(11) &  0.147(7) & 0.73(13) \\ \hline
$f_{\rm NE2571A}^{\rm Clinac}$ & 0.85(6) & 0.12(4) & 0.4(5) \\
$f_{\rm NE2571A}^{\rm mono}$ & 0.80(6) & 0.17(6) & 0.3(2)\\ \hline
$f_{\rm NE2581A}^{\rm Clinac}$ & 0.85(2) & 0.127(9) & 0.6(2) \\
$f_{\rm NE2581A}^{\rm mono}$ & 0.74(7) & 0.23(6) & 0.23(10) \\ 
\hline\hline
\end{tabular}
\end{center}
\end{table}

In the case of the NE2571 chamber (figure \ref{fig:UNO}a), our results are compared to those found by Muir and Rogers \cite{Muir13} who provided $k_{{\rm Q,Q}_0}$ values for clinical (solid circles) and monoenergetic (open circles) beams. Also their fit
\begin{equation}
h_{\rm NE2571}\, = \, 0.8823\,+\,0.1042\,x^{-0.7928} 
\label{eq:fitMR}
\end{equation}
has been included with a dotted line.

\begin{figure}[!th]
\centering
\includegraphics[width=6cm]{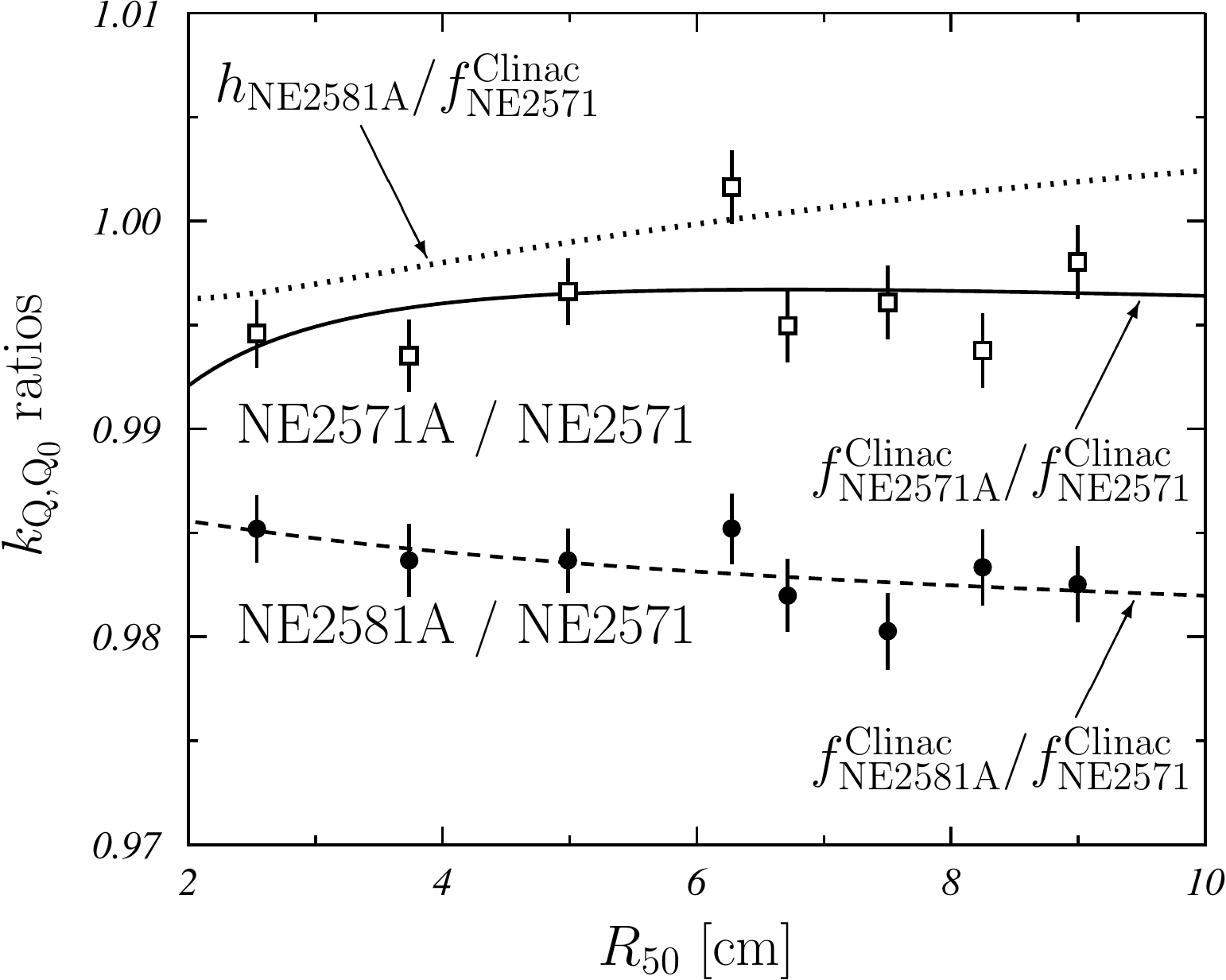}
\caption{Ratios of the $k_{{\rm Q,Q}_0}$ values obtained for the NE2571A (open squares) and NE2581A (solid circles) to those of the NE2571. The solid and dashed curves represent, respectively, the ratios  $f_{\rm NE2571A}^{\rm Clinac} / f_{\rm NE2571}^{\rm Clinac}$ and  $f_{\rm NE2581A}^{\rm Clinac} / f_{\rm NE2571}^{\rm Clinac}$ between the corresponding fitting functions defined in equation (\ref{eq:fit}). The dotted curve shows the ratio between the fitting function $h_{\rm NE2571}$ obtained by Muir and Rogers for the NE2571 chamber \cite{Muir13}, given in equation (\ref{eq:fitMR}), and our $f_{\rm NE2571}^{\rm Clinac}$.
\label{fig:DOS}
}
\end{figure}

In figure \ref{fig:DOS} the $k_{{\rm Q,Q}_0}$ values obtained for the NE2571A and NE2581A chambers are compared to those found for the NE2571 by means of the corresponding ratios that are shown with open squares and solid circles, respectively. In addition, the ratios of the fitting functions are also given by the solid and dashed curves. For comparison, the ratio between the fitting function $h_{\rm NE2571}$, defined in equation (\ref{eq:fitMR}), quoted by Muir and Rogers \cite{Muir13}, and the $f_{\rm NE2571}^{\rm Clinac}$ we have obtained is shown by the dotted curve.

Several facts deserve a comment. First, the $k_{{\rm Q,Q}_0}$ values we have obtained for the NE2571 agree with those quoted by Muir and Rogers \cite{Muir13}. By comparing our results with those resulting from a quadratic interpolation of the $k_{{\rm Q,Q}_0}$ found by Muir and Rogers (in order to have the same $R_{50}$ values) the largest difference in the case of the clinical beams was below 0.2\%. Larger discrepancies were found for the monoenergetic beams, the maximum difference being 1.1\% for the beam with the biggest energy which, on the other hand, is out of the range covered by the data of Muir and Rogers that reached up to $R_{50}=8.12\,$cm. This overall agreement in corroborated by the dotted curve in figure \ref{fig:DOS}. In the $R_{50}$ range analyzed, the relative differences between the respective fits to the clinical beams data vary from 0.4\% to 0.2\% as $R_{50}$ increases.

The second point to be noted is linked to the differences observed between the calculations performed with the complete Clinac beams and the monoenergetic beams reproducing the corresponding $R_{50}$ values. The results we obtained are similar for the three ionization chambers studied as shown in figure \ref{fig:UNO} and in table \ref{tab:DOS}. Monoenergetic $k_{{\rm Q,Q}_0}$ values overestimate by $\sim 0.4\%$ those found for the Clinac beams at low $R_{50}$ while the contrary occurs for $R_{50}$ above 6-$7\,$cm where the differences are slightly larger than $0.5\%$. A similar behavior was pointed for the Exradin A10, A11, A11TW, P11, P11TW, T11 and T11TW, manufactured by Standard Imaging (Middleton, USA), as well as for the NACP-02, all of them plane-parallel ionization chambers \cite{Erazo14}. In this previous work it was demonstrated that an important contribution to this discrepancy may be ascribed to the secondary photons produced in the linac head and the air separating it from the phantom.

It is also worth pointing out the systematic differences we obtained between the $k_{{\rm Q,Q}_0}$ values for the NE2571 and the NE2571A, which  is supposed to substitute the former. Figure \ref{fig:DOS} shows the importance of these differences. Except for the anomalous data point around $R_{50} \sim 6\,$cm, the $k_{{\rm Q,Q}_0}$ ratio remains significantly below 1, with an average value of $\sim 0.995$ (see open squares). This result is also obtained when the ratio $f_{\rm NE2571A}^{\rm Clinac} / f_{\rm NE2571}^{\rm Clinac}$ (solid curve) is considered. This indicates that the NE2571A $k_{{\rm Q,Q}_0}$ values underestimate those of its predecessor chamber, the NE2571, by 0.5\% roughly.

Finally, also in figure \ref{fig:DOS} it is apparent that the NE2581A $k_{{\rm Q,Q}_0}$ values are systematically below those of the NE2571 by $\sim 1.6\%$ as the $k_{{\rm Q,Q}_0}$ ratios (solid circles) and the ratio $f_{\rm NE2581A}^{\rm Clinac} / f_{\rm NE2571}^{\rm Clinac}$ (dashed curve) indicate.

\section*{Conclusions}

In this work the Monte Carlo code PENELOPE was used to calculate the beam quality correction factor, $k_{Q,Q_{0}}$ for three ionization chambers (NE2571, NE2571A and NE2581A) in case of electron beam dosimetry. Eight Varian Clinac 2100 C/D electron beams, with nominal energies between 6 and $22\,$MeV, were simulated using the geometries generated with the code {\sc penEasyLinac}. For these beams and chambers, the absorbed dose in water and in the air cavity of the real detectors were determined and this permitted to calculate the overall perturbation factors $f_{\rm c,Q}$ and the corresponding $k_{{\rm Q,Q}_0}$ by dividing the $f_{\rm c,Q}$ factors with the reference value obtained for a $^{60}$Co beam. These $k_{{\rm Q,Q}_0}$ factors were also obtained in a similar way for monoenergetic beams that were tuned to reproduce the $R_{50}$ quality index of the Clinac beams.

The results we have obtained for the NE2571 ionization chamber are in good agreement with those quoted by Muir and Rogers who did their simulations with the EGSnrc Monte Carlo code. The differences observed are well below 0.5\%.

The $k_{{\rm Q,Q}_0}$ values obtained for the monoenergetic beams are larger than those ones calculated for the Clinac ones at low $R_{50}$ values and the contrary occurs above $6\,$cm, with differences that may be slightly larger than $0.5\%$. This behavior appears to be similar to that found for other plane-parallel ionization chambers in previous works.

The $k_{{\rm Q,Q}_0}$ factors found for the NE2571A chamber show a systematic difference with those of its predecessor chamber, the NE2571, the latter being $\sim 0.5\%$ larger than the former.


\subsection*{Acknowledgements}
This work has
been supported in part by the Junta de Andaluc\'{\i}a (FQM0387), by
the Ministerio de Econom\'{\i}a y Competitividad (FPA2015-67694-P) and by the European
Regional Development Fund (ERDF). LB acknowledges financial support from the Deutsche Forschungsgemeinschaft project BR 4043/1-1.

\end{document}